\documentclass{INTERSPEECH2023}

\usepackage{url}
\usepackage{cite}
\usepackage{multirow}
\usepackage{balance} 
\usepackage{footmisc}

\newcommand{\ReduceSpaceUnderFigure}{\vspace{-0pt}}
\newcommand{\ReduceSpaceUnderTable}{\vspace{-1pt}}

\interspeechcameraready

\title{LibriTTS-R: A Restored Multi-Speaker Text-to-Speech Corpus}
\name{
Yuma~Koizumi$^1$,
Heiga~Zen$^1$,
Shigeki~Karita$^1$,
Yifan~Ding$^1$,
Kohei~Yatabe$^2$,
Nobuyuki~Morioka$^1$,\\
Michiel~Bacchiani$^1$,
Yu~Zhang$^3$,
Wei~Han$^3$,
Ankur~Bapna$^3$
}
\address{$^1\,$Google, Japan, $^2\,$Tokyo University of Agriculture \& Technology, Japan, $^3\,$Google, USA}
\email{\{koizumiyuma,heigazen,karita\}@google.com}

\begin{document}

\maketitle
 
\begin{abstract}
This paper introduces a new speech dataset called ``LibriTTS-R'' designed for text-to-speech (TTS) use. 
It is derived by applying speech restoration to the LibriTTS corpus, which consists of 585 hours of speech data at 24\,kHz sampling rate from 2,456 speakers and the corresponding texts.
The constituent samples of LibriTTS-R are identical to those of LibriTTS, with only the sound quality improved.
Experimental results show that the LibriTTS-R ground-truth samples showed significantly improved sound quality compared to those in LibriTTS. In addition, neural end-to-end TTS trained with LibriTTS-R achieved speech naturalness on par with that of the ground-truth samples. The corpus is freely available for download from \url{http://www.openslr.org/141/}.
\end{abstract}

\section{Introduction}
\label{sec:intro}

Text-to-speech (TTS) technologies have been rapidly advanced along with the development of deep learning~\cite{wavenet,tacotron2,pngbert,parallel_tacotron,fastspeech,fastspeech2}.
With studio-quality recorded speech data, one can train acoustic models~\cite{pngbert,tacotron2} and high-fidelity neural vocoders~\cite{Kong_2020,wavefit}. These have enabled us to synthesize speech in a reading style almost as natural as human speech.
In addition, many implementations of the latest TTS models have been published~\cite{espnet_tts,speechbrain}, and the gateway to TTS research is certainly widening.

One of the remaining barriers to develop high-quality TTS systems is the lack of large and high-quality public dataset.
Training of high-quality TTS models requires a large amount of studio-quality data.
In several TTS papers, over 100 hours of studio-recorded data have been used~\cite{nat,pngbert,wavefit}.
Unfortunately, such studio-recorded datasets are not publicly available, and thus reproducing their results is difficult for others.

At the same time, speech restoration (SR) has advanced using speech generative models~\cite{Maiti_waspaa_2019,Maiti_icassp_2020,self_remaster,Su_2020,Su_2021,voice_filxer,UNIVERSE}.
These state-of-the-art models can convert reverberated lecture and historical speech to studio-recorded quality~\cite{Su_2021,voice_filxer,UNIVERSE}.
Inspired by these results, we came up with an idea that the above-mentioned barrier can be removed by applying SR to public datasets.

With this paper, we publish \textit{LibriTTS-R}, a quality-improved version of LibriTTS~\cite{libritts}.
LibriTTS is a non-restrictive license multi-speaker TTS corpus consisting of 585 hours of speech data from 2,456 speakers and the corresponding texts.
We cleaned LibriTTS by applying a text-informed SR model, \textit{Miipher},~\cite{miipher} that uses w2v-BERT~\cite{w2vbert} feature cleaner and WaveFit neural vocoder~\cite{wavefit}.
By subjective experiments, we show that the speech naturalness of a TTS model trained with LibriTTS-R is greatly improved from that trained with LibriTTS, and is comparable with that of the ground-truth.

LibriTTS-R is publicly available at \url{http://www.openslr.org/141/}, with the same non-restrictive license.
Audio samples of the ground-truth and TTS generated samples are available at our demo page\footnote{\url{https://google.github.io/df-conformer/librittsr/}\label{footnote:demo}}.

\section{The LibriTTS corpus}
\label{sec:libritts_sec}

The LibriTTS corpus is one of the largest multi-speaker speech datasets designed for TTS use. This dataset consists of 585 hours of speech data at 24 kHz sampling rate from 2,456 speakers and the corresponding texts. The audio and text materials are derived from the LibriSpeech corpus~\cite{librispeech}, which has been used for training and evaluating automatic speech recognition systems. Since the original LibriSpeech corpus has several undesired properties for TTS including sampling rate and text normalization issues, the samples in LibriTTS were re-derived from the original materials (MP3 from LibriVox and texts from Project Gutenberg) of LibriSpeech.

One issue is that the LibriTTS sound quality is not on par with smaller scale but higher quality TTS datasets such as LJspeech~\cite{ljspeech17}. The quality of the TTS output is highly affected by that of the speech samples used in model training. Therefore, the quality of the generated samples of a TTS model trained on LibriTTS doesn't match those of the ground-truth samples~\cite{glowtts,valle2021flowtron}.
For example, Glow-TTS achieved 3.45 mean-opinion-score (MOS) on LibriTTS where the speech obtained from the ground-truth mel-spectrograms by a vocoder was 4.22~\cite{glowtts}. Note that MOSs on the LJspeech for generated and ground-truth were 4.01 and 4.19, respectively~\cite{glowtts}.
The results suggest that the quality of speech samples in LibriTTS are inadequate for training of high-quality TTS models.

\section{Data processing pipeline}
\label{sec:speech_restoration}

Although noisy TTS datasets are useful for advanced TTS model training~\cite{yourtts,valle,spear_tts}, access to large scale high-quality datasets is as equally important for advancing TTS techniques. To provide a public large-scale and high-quality TTS dataset, we apply a SR model to LibriTTS.

\begin{figure}[t]
  \centering
\includegraphics[width=0.8\linewidth,clip]{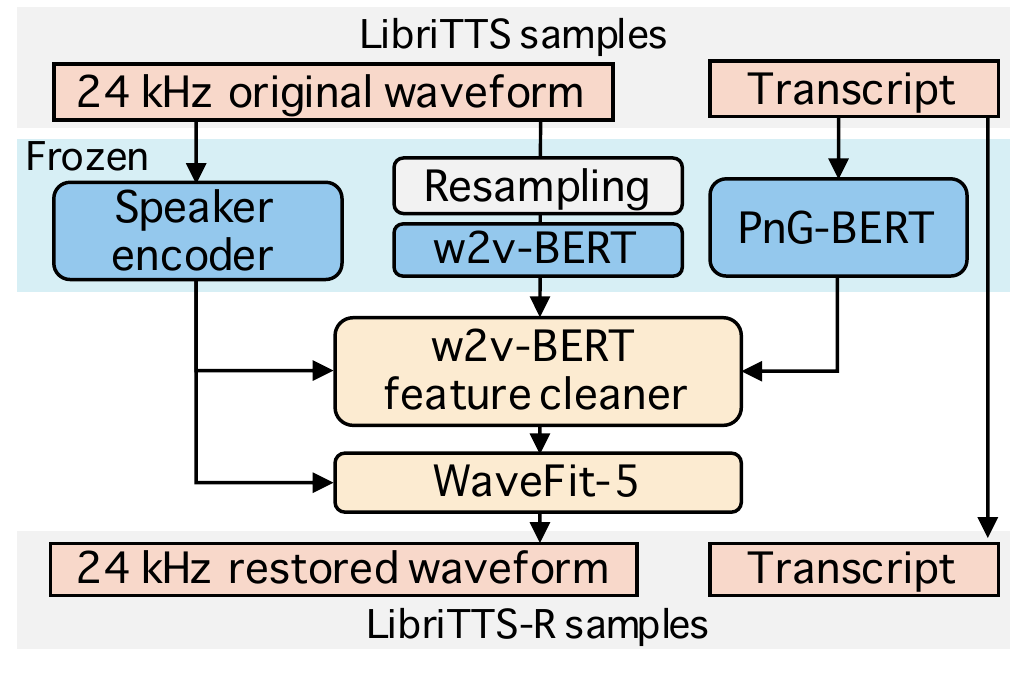} 
  \vspace{-8pt}
  \caption{Data processing pipeline overview. Speech samples in the LibriTTS corpus are restored using Miipher~\cite{miipher}.}
  \label{fig:model_overview}
  \ReduceSpaceUnderFigure
\end{figure}

\subsection{Speech restoration model overview}
\label{sec:model_overview}

One critical requirement of SR models for the purpose of cleaning datasets is robustness.
If the SR model generates a large number of samples with artifacts, it will adversely impact the subsequent TTS model training.
Therefore, for our purposes, we need to reduce as much as possible the number of samples that fail to be recovered.

To satisfy this requirement, we use a text-informed parametric re-synthesis-based SR model, \textit{Miipher}~\cite{miipher}, as shown in Fig.~\ref{fig:model_overview}.
In this model, first, w2v-BERT features are extracted by w2v-BERT~\cite{w2vbert} from the noisy waveform. Then, a DF-Conformer~\cite{Koizumi_waspaa_2021}-based feature-cleaner predicts the w2v-BERT features of the clean waveform. Finally, the restored waveform is synthesized using a WaveFit-5 neural vocoder~\cite{wavefit}.

The reason for selecting Miipher is that it adresses two particularly difficult to restore degradation patterns observed in LibriTTS samples.
The first degradation is phoneme masking. Speech signals are sometimes masked by noise and/or reverberation, resulting in speech that is is difficult to discriminate from noise without additional information.
The second degradation is phoneme deletion. Important frequency parts of some phonemes could be missing from the signal due to non-linear audio processing and/or down-sampling.
To address these problems, Miipher introduced two techniques.
(i) for the input feature, it uses w2v-BERT~\cite{w2vbert} features instead of log-mel spectrogram used in a conventional SR model~\cite{voice_filxer},
and
(ii) to use linguistic features conditioning extracted by PnG-BERT~\cite{pngbert} from the transcript corresponding to the noisy speech.
Since w2v-BERT is trained on large amounts of degraded speech samples and it improves ASR performance, we expect its effectiveness on making SR models robust against speech degradation.
In addition, the use of text information improving speech inpainting performance~\cite{borsos22_interspeech}, we consider that it also improves speech restoration performance.
For the detail, please see the original paper~\cite{miipher}.

\subsection{Speech restoration model training}
\label{sec:model_traiing}

We trained a Miipher model with a proprietary dataset that contains 2,680 hours of noisy and studio-quality speech pairs.
The target speech dataset contains 670 hours of studio-recorded Australian, British, Indian, Nigerian, and North American English at 24 kHz sampling. For the noise dataset, we used the TAU Urban Audio-Visual Scenes 2021 dataset~\cite{tau_2021_dataset}, internally collected noise snippets that simulate conditions like cafe, kitchen, and cars, and noise sources.
The noisy utterances were generated by mixing randomly selected speech and noise samples from these datasets with signal-to-noise ratio (SNR) from 5 dB to 30 dB.
In addition, we augmented the noisy dataset with 4 patterns depending on the presence or absence of reverberation and codec artifacts.
A room impulse response (RIR) for each sample was generated by a stochastic RIR generator using the image method~\cite{image_method}. For simulating codec artifacts, we randomly applied one of MP3, Vorbis, A-law, Adaptive Multi-Rate Wideband (AMR-WB), and OPUS with a random bit-rate.
The detailed simulation parameters were listed in~\cite{miipher}.

We first pre-trained the feature-cleaner and WaveFit neural vocoder 150k and 1M steps, respectively, where WaveFit was trained to reconstruct waveform from clean w2v-BERT features. Then, we fine-tuned the WaveFit neural vocoder 350k steps using cleaned w2v-BERT features by the pre-trained feature-cleaner.

\subsection{Speech restoration pipeline}
\label{sec:pipeline}

First, we calculated PnG-BERT~\cite{pngbert} features from a transcript and
a speaker embedding using the speaker encoder described in~\cite{miipher} from the original 24 kHz sampling waveform.
Here, for speech samples with waveform lengths shorter than 2 seconds, the speaker embedding was calculated after repeating them to get a pseudo longer waveform.
Since the w2v-BERT~\cite{w2vbert} model was trained on 16 kHz waveforms, we applied down-sampling to the LibriTTS sample for calculating w2v-BERT features.
Finally, we synthesized restored 24 kHz sampling waveform using WaveFit~\cite{wavefit}.

\section{Experiments}
\label{sec:experiment}

\subsection{Subjective experiments for ground-truth samples}
\label{sec:mos_gt_samples}

\subsubsection{Experimental setups}

We first compared the quality of ground-truth speech samples in LibriTTS-R with those in LibriTTS.
We evaluated the sound quality using ``test-clean'' and ``test-other'' subsets. We randomly selected 620 samples from each subset. Since the ``train-*'' and ``dev-*'' subsets are also divided into ``clean'' and ``other'' according to the same word-error-rate (WER)-based criteria, the sound quality of the entire dataset can be predicted by evaluating the sound quality of these two subsets.

To evaluate subjective quality, we rated speech quality through mean-opinion-score (MOS) and side-by-side (SxS) preference tests.
We asked to rate the naturalness in MOS test, and ``\textit{which sound quality is better?}'' in SxS test.
The scale of MOS was a 5-point scale (1:~Bad, 2:~Poor, 3:~Fair, 4:~Good, 5:~Excellent) with rating increments of 0.5, and that of SxS was a 7-point scale (-3 to 3).
Test stimuli were randomly chosen and each stimulus was evaluated by one subject. Each subject was allowed to evaluate up to six stimuli, that is, over 100 subjects participated in this experiment to evaluate 640 samples in each condition. The subjects were paid native English speakers in the United States. They were requested to use headphones in a quiet room.
Audio samples are available in our demo page~\footref{footnote:demo}.

\subsubsection{Results}

\begin{figure*}[t]
\vspace{-1pt}
  \centering
  \includegraphics[width=\linewidth,clip]{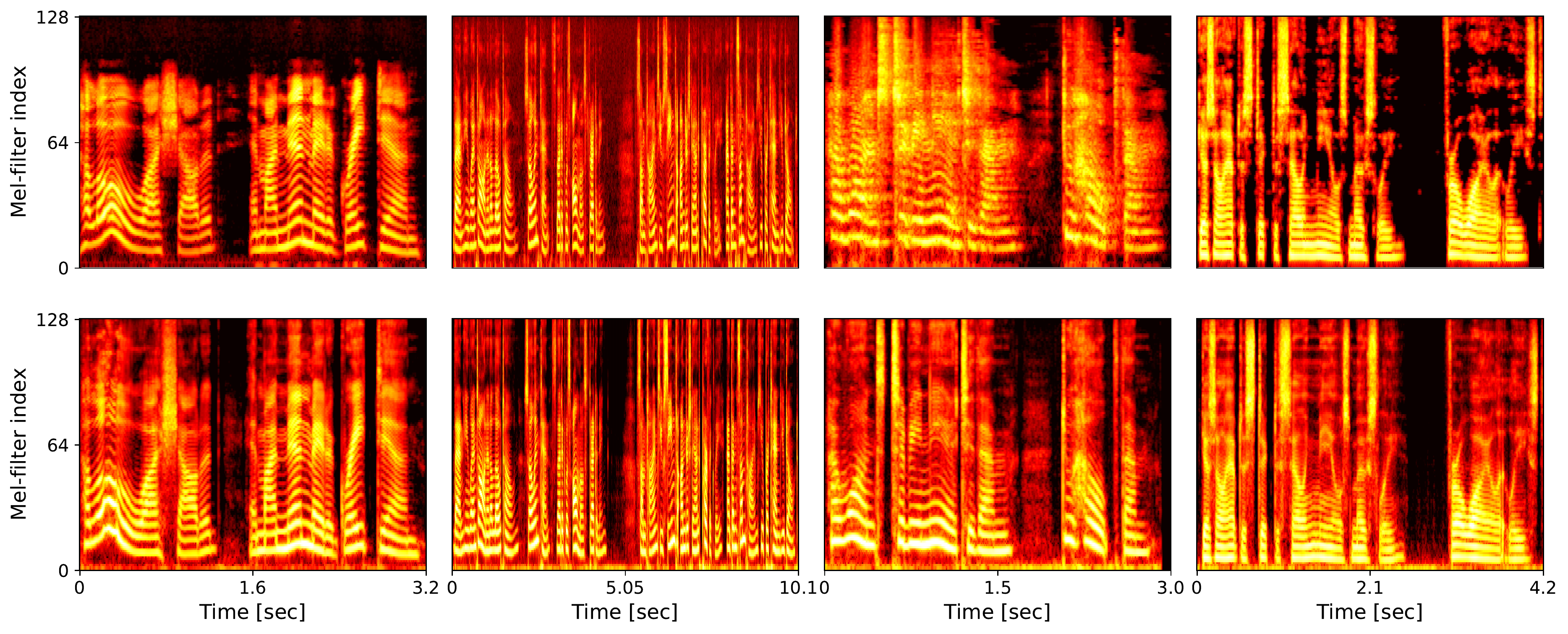} 
  \caption{Log-mel spectrograms of ground-truth waveforms from (top) LibriTTS and (bottom) LibriTTS-R. The left two and right two examples are from ``test-clean'' and ``test-other'' splits, respectively.}
  \label{fig:gt_example}
  \ReduceSpaceUnderFigure
\end{figure*}

\begin{table}[ttt]
\caption{MOS and SxS test results on the ground-truth samples with their 95\% confidence intervals. A positive SxS score indicates that LibriTTS-R was preferred.}
\label{tab:mos_sxs_gt_result}
\centering
\begin{tabular}{ c | c c | c}
\toprule
\multirow{2}{*}{\textbf{Split}} & \multicolumn{2}{c|}{\textbf{MOS} ($\uparrow$)} & \multirow{2}{*}{\textbf{SxS}} \\	
 & LibriTTS & LibriTTS-R & \\	
\midrule
test-clean & $4.36 \pm 0.08$ & $\bm{4.41 \pm 0.07}$  & $0.80 \pm 0.15$ \\
test-other & $3.94 \pm 0.10$ & $\bm{4.09 \pm 0.10}$  & $1.42 \pm 0.14$ \\
\bottomrule
\end{tabular}
\ReduceSpaceUnderTable
\end{table}

Table~\ref{tab:mos_sxs_gt_result} shows the MOS and SxS test results.
In terms of speech naturalness, LibriTTS achieved high MOSs: 4.36 and 3.94 on test-clean and test-other, respectively. Although LibriTTS-R achieved better MOSs than LibriTTS in both splits, the difference was not significant. The reason of small difference in naturalness might be because ground-truth samples are real speech spoken by humans. In contrast, in terms of sound quality rated by SxS tests, significant differences were observed on both split. 

To confirm whether the text-content and speaker in the restored speech samples are maintained, we evaluated the WER and speaker similarity.
To compute WER, we used ``Pre-trained Conformer XXL'' model proposed in \cite{yu_asr_2020}. WER of ``test-clean'' and ``test-other'' splits of LibriTTS were $3.4$ and $5.1$, whereas those of LibriTTS-R were $3.2$ and $5.1$, respectively\footnote{WER were a bit higher than those reported in the original paper~\cite{yu_asr_2020}, because the ASR model was trained on noisy speech and transcripts normalized by a different text-normalizer.}.
Therefore, the text contents are considered to be not changed.
To evaluate speaker similarity, we used the cosine similarity of speaker embedding~\cite{NEURIPS2018_6832a7b2,chen2018sample}. We calculated the similarity between the different utterances spoken by the same speaker in the same dataset.
This is because the samples in LibriTTS are distorted, even if the similarity between corresponding samples in LibriTTS and LibriTTS-R is small, this does not necessarily indicate speaker similarity.
The cosine similarity of LibriTTS ``test-clean'' and ``test-other'' splits were $0.784$ and $0.755$, and those of LibriTTS-R were $0.762$ and $0.745$.
Since the similarity calculated from the samples in LibriTTS spoken by different speakers was $0.302$, the speech characteristics of each speaker is considered to be consistent.

Figure~\ref{fig:gt_example} shows the 128-dim log-mel spectrogram of speech samples from LibriTTS and LibriTTS-R.
We can see that the LibriTTS samples are degraded by a variety of factors even if these are from the test-clean split: from left to right, it can be considered that speech samples were degraded by down-sampling, environmental noise, reverberation, and non-linear speech enhancement, respectively.
As we can see spectrograms of LibriTTS-R samples, the SR model restored these speech samples into high-quality ones.
This could be the reason of the significant differences in the SxS tests.

Note that we have found a few examples that LibriTTS speech sample achieved a better score in SxS comparison. By listening these examples, two of 640 LibriTTS-R speech samples were distorted due to the failure of SR.
Since it is difficult to manually check all samples, we have not checked all speech samples in LibriTTS-R. Therefore, the samples in training splits may also contain a small number of distorted samples.

\subsection{Subjective experiments for TTS generated samples}
\label{sec:mos_tts_samples}

\begin{figure*}[t]
\vspace{-1pt}
  \centering
  \includegraphics[width=\linewidth,clip]{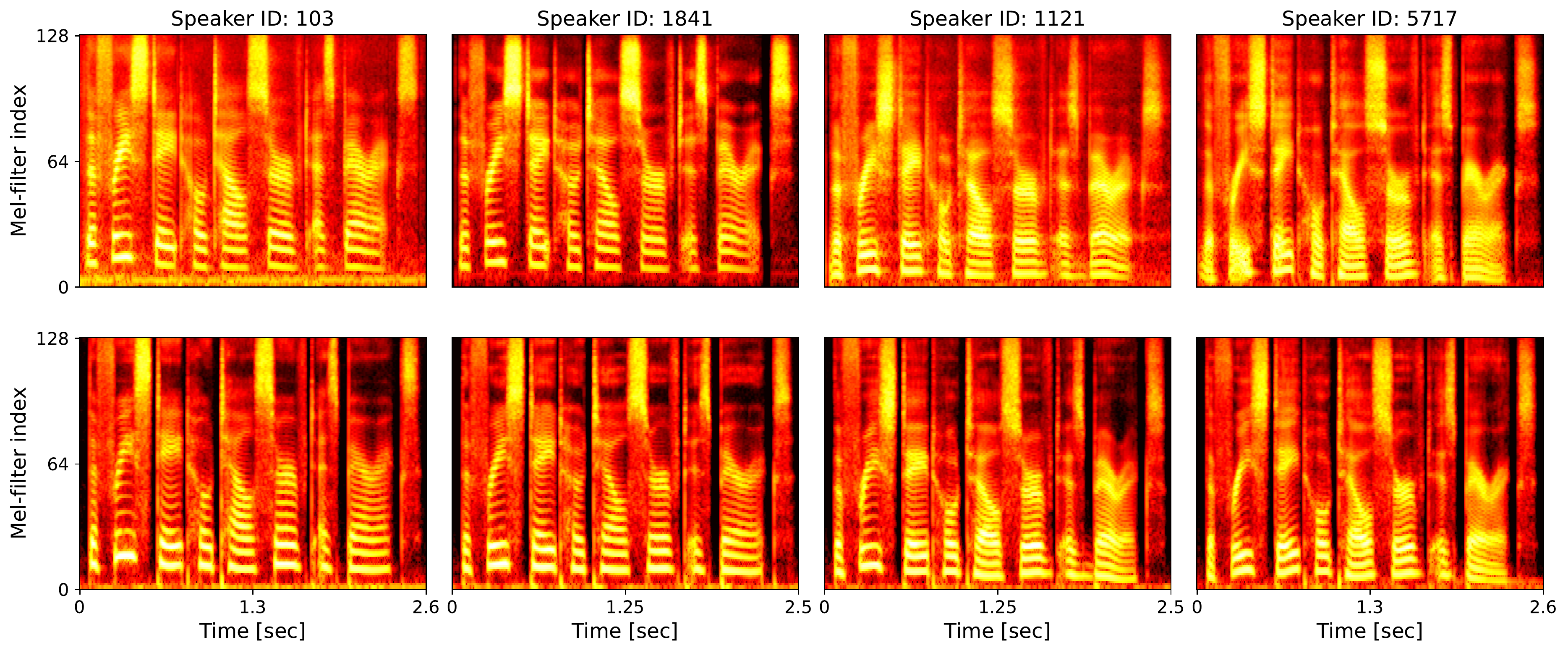} 
  \caption{Log-mel spectrograms of TTS generated waveforms where the multi-speaker TTS model was trained on (top) LibriTTS and (bottom) LibriTTS-R, respectively. The input text was ``The Free State Hotel served as barracks''.}
  \label{fig:tts_example}
  \ReduceSpaceUnderFigure
\end{figure*}

\begin{table*}[ttt]
\caption{MOSs for the baseline multi-speaker TTS model outputs with their 95\% confidence intervals.}
\label{tab:mos_result}
\centering
\begin{tabular}{ c | c c c | c c c }
\toprule
\multirow{2}{*}{\textbf{Training dataset}} & \multicolumn{6}{c}{\textbf{Speaker ID}} \\	
& 19 & 103 & 1841 & 204 & 1121 & 5717 \\
\midrule
LibriTTS Train-460  & $2.49 \pm 0.10$ & $2.94 \pm 0.10$ & $3.40 \pm 0.09$ & $2.88 \pm 0.10$ & $2.72 \pm 0.10$ & $2.86 \pm 0.09$ \\
LibriTTS Train-960  & $2.59 \pm 0.10$ & $2.75 \pm 0.10$ & $3.35 \pm 0.10$ & $2.74 \pm 0.09$ & $2.83 \pm 0.10$ & $2.97 \pm 0.10$ \\
\midrule
LibriTTS-R Train-460  & $\bm{4.11 \pm 0.08}$ & $4.09 \pm 0.08$ & $3.88 \pm 0.09$ & $3.67 \pm 0.09$ & $3.92 \pm 0.09$ & $3.67 \pm 0.08$ \\
LibriTTS-R Train-960  & $4.06 \pm 0.08$ & $\bm{4.31 \pm 0.08}$ & $\bm{4.20 \pm 0.08}$ & $\bm{4.11 \pm 0.08}$ & $\bm{4.23 \pm 0.07}$ & $\bm{4.08 \pm 0.08}$ \\
\bottomrule
\end{tabular}
\ReduceSpaceUnderTable
\end{table*}

\begin{table}[ttt]
\caption{SxS test results on the baseline multi-speaker TTS model outputs with their 95\% confidence intervals. A positive score indicates that training on LibriTTS-R was preferred.}
\label{tab:sxs_result}
\centering
\begin{tabular}{ c | c c}
\toprule
\multirow{2}{*}{\textbf{Speaker ID}} & \multicolumn{2}{c}{\textbf{Training dataset}} \\	
  & Train-460  & Train-960 \\	
\midrule
19  & $2.38 \pm 0.11$ & $2.51 \pm 0.10$ \\	
204 & $1.84 \pm 0.14$ & $2.20 \pm 0.12$ \\	
\bottomrule
\end{tabular}
\ReduceSpaceUnderTable
\end{table}

\subsubsection{Experimental setups}

We trained multi-speaker TTS models with the same architecture and the same hyper-parameters using either the LibriTTS or LibriTTS-R corpus.
The TTS model was build by concatenating the following acoustic model and neural vocoder without joint fine-tuning.\\
\vspace{2pt}
\noindent
\textbf{Acoustic model:} We used a duration unsupervised Non-Attentive Tacotron (NAT) with a fine-grained variational auto-encoder (FVAE)~\cite{nat}. We used the same hyper-parameters and training parameters listed in the original paper~\cite{nat}.
We trained this model for 150k steps with a batch size of 1,024.\\
\vspace{2pt}
\noindent
\textbf{Neural vocoder:} We used a WaveRNN~\cite{wavernn} which consisted of a single long short-term memory layer with 512 hidden units, 5 convolutional layers with 512 channels as the conditioning stack to process the mel-spectrogram features, and a 10-component mixture of logistic distributions as its output layer.
The learning rate was linearly increased to $10^{-4}$ in the first 100 steps then exponentially decayed to $10^{-6}$ from 200k to 300k steps.
We trained this model using the Adam optimizer~\cite{kingma2014adam} for 500k steps with a batch size of 512.

The TTS model was trained on two types of training datasets: Train-460 and Train-960.
Train-460 consists of the ``train-clean-100'' and ``train-clean-360'' subsets, and Train-960 indicates using ``train-other-500'' in addition to Train-460.

For the test sentences, we randomly selected 620 evaluation sentences from the test-clean split. We synthesized waveforms with 6 speakers (three female and three male) those are used in the LibriTTS baseline experiments~\cite{libritts}. The female and male reader IDs were (19, 103, 1841) and (204, 1121, 5717), respectively. 
To evaluate subjective quality, we rated speech naturalness through MOS and side-by-side (SxS) preference tests. The listening test setting was the same as Sec.~\ref{sec:mos_gt_samples}
Audio samples of generated speech are available in our demo page~\footref{footnote:demo}.

\subsubsection{Results}

Table~\ref{tab:mos_result} shows the MOS results.
In all speaker IDs except for ID 19, the TTS model using LibriTTS-R Train-960 as the training dataset achieved the highest MOSs. For Speaker ID 19, the model using LibriTTS-R Train-460 achieved the highest MOS, which was not significantly different from that using LibriTTS-R Train-960. In other speaker IDs, MOSs of LibriTTS-R Train-960 were significantly better than that of LibriTTS-R Train-460. This trend was not observed in LibriTTS, rather in some cases, MOS was decreased by using LibriTTS Train-960. The reason for this degradation might be because that the ``train-other-500'' split contains a lot of degraded speech samples. This result suggests that the use of LibriTTS ``train-other-500'' split rather degrades the output quality of the TTS. In contrast, speech samples in LibriTTS-R ``train-other-500'' split are restored to high-quality speech samples, and resulting in that enables us to use a large amount of high-quality training data and improved the naturalness of the TTS outputs.
In addition, the TTS model trained on LibriTTS-R Train-960 achieved MOSs on a par with human spoken speech samples in LibriTTS, effects of a few distorted speech samples in the training can be considered as not significant.

Table~\ref{tab:sxs_result} shows the SxS results. We observed that the use of LibriTTS-R also improve not only naturalness but also the sound quality of TTS outputs.
Figure~\ref{fig:tts_example} shows 128-dim log-mel spectrograms of TTS outputs. We can see the harmonic structure is broken in the ID 5717 output of the TTS model trained on LibriTTS (top right). The presence of such a sample could be the reason for the lower naturalness scores on the MOS test.
Also, from ID 103 and 1121 examples, we can observe background noise in the output of TTS model trained on LibriTTS. Such background noise does not exist in the outputs of TTS model trained on LibriTTS-R.
From these results, we conclude that the LibriTTS-R corpus is a better TTS corpus that the LibriTTS corpus, and enables us to train a high-quality TTS model. 

\section{Conclusions}
\label{sec:conclusions}

This paper introduced LibriTTS-R, a sound quality improved version of LibriTTS~\cite{libritts}.
We cleaned speech samples in the LibriTTS corpus by applying an SR model~\cite{miipher}.
By subjective experiments, we show that the speech naturalness of a TTS model trained with LibriTTS-R is improved from that trained with LibriTTS, and is comparable with that of the ground-truth. This corpus is released online, and it is freely available for download from \url{http://www.openslr.org/141/}. We hope that the release of this corpus accelerates TTS research.

\clearpage 
\balance
\bibliographystyle{IEEEtran}
\bibliography{mybib}

\end{document}